\documentclass[aps,prl,groupedaddress,showpacs,tighten,nofootinbib,preprint]
{revtex4}

\usepackage{graphicx}

\usepackage{times}

\begin{document}

\preprint{FERMILAB-PUB-06-022-A}

\title{Pierre Auger Data, Photons, and Top-Down Cosmic Ray Models}

\author{Nicolas Busca,$^{1,2,3}$ Dan Hooper,$^1$ and Edward W.~Kolb$^{1,3}$}

\affiliation{$^1$Fermi National Accelerator Laboratory, 
	Particle Astrophysics Center, Batavia, IL  60510-0500}
	
\affiliation{$^2$ Kavli Institute of Cosmological Physics, 
	The University of Chicago, Chicago, IL~~60637-1433}
	
\affiliation{$^3$Department of Astronomy \& Astrophysics,
	The University of Chicago, Chicago, IL~~60637-1433}

\date{\today}

\begin{abstract}

We consider the ultra-high energy cosmic ray (UHECR) spectrum
as measured by the Pierre Auger Observatory. Top-down models for the origin of
UHECRs predict an increasing photon component at energies above about
$10^{19.7}$eV. Here we present a simple prescription to compare the Auger data
with a prediction assuming a pure proton component or a prediction assuming
a changing primary component appropriate for a top-down model. We find that the
UHECR spectrum predicted in top-down models is a good fit to the Auger data.
Eventually, Auger will measure a composition-independent spectrum and will be
capable of either confirming or excluding the quantity of photons predicted in
top-down models.

\end{abstract}

\pacs{95.55.Vj;96.50.sb;98.70.Sa\hspace{0.5cm} FERMILAB-PUB-06-022-A}

\maketitle

\section{Introduction}

The origin of the highest energy cosmic rays has been a subject of great
interest for some time. The particular aspect of Ultra-High Energy Cosmic Ray
(UHECR) physics that has raised the greatest degree of interest is the
question of whether the cosmic ray spectrum exhibits a feature known as the GZK
cutoff \cite{gzk}. This cutoff is the result of the suppression of the cosmic
ray spectrum above a few $10^{19}$ eV due to protons interacting with the cosmic
microwave background. If a GZK suppression is not observed, it would indicate
that the sources of these ultra-high energy events would be of local origin,
cosmologically speaking (within 10 to 50 Mpc). However, no nearby astrophysical
sources sources capable of accelerating particles to such high energies are
known to exist.

Measurements of the UHECR spectrum have not clearly settled the issue of
whether a GZK suppression is present. On one hand, the spectrum  measured by
the AGASA experiment shows no indication of a GZK suppression \cite{ag}. In
particular, in Ref.\ \cite{agasa} it is shown that the number of events with
energies above $10^{20}$eV expected in a GZK scenario is 3.6, while the number
of events observed by AGASA is 11, corresponding to a significance of 3 standard
deviations. In contrast, the HiRes experiment appears to have observed the
presence of a GZK suppression: Ref.\ \cite{hires} concludes that HiRes data
exclude a non-GZK scenario with a significance of 3 to 4 standard deviations.
Given this  discrepancy, it appears that further data would be required to
resolve the question at hand. In particular, the Pierre  Auger Observatory (or
simply Auger) is currently under construction at its southern site in
Argentina. Auger combines  the techniques used by AGASA  (an air shower ground
array) and HiRes (fluorescence detectors) allowing it to make energy spectrum
measurements which are less composition and model dependent than either HiRes
or AGASA.
                                                                                                                             
Auger's first results were released in 2005, but did not clearly resolve the
question of whether a GZK feature is present  in the UHECR spectrum. These data
was collected with only a fraction of the southern site completed, but yet the
total  exposure at this point was slightly larger than the total accumulated by
AGASA. Auger calibrated their ground array data using their florescence
detectors, resulting in a largely composition independent energy measurement.
Due to a lack of  events, this hybrid calibration was only possible at energies
well below the GZK cutoff, however, and if the composition  of the UHECR
spectrum charges between the calibration energies and higher energies, then the
highest energy bins in the  published Auger spectrum must be modified.
                                                                                                                             
This effect is particularly pronounced in the case of photon primaries. Auger's
ground array measures energy with a  parameter known as S(1000) \cite{augerspec}, which is
proportional to the water Cherenkov signal in the surface array at a distance
of 1000 meters from  the shower axis. Due to the lack of muons generated by
photon primaries, photon-induced events would produce a smaller S(1000) than
proton-induced events of the same primary energy. Therefore, the actual cosmic
ray flux may be considerably larger at the highest energies than reported by
Auger if a substantial fraction of the highest energy cosmic rays are photons.

A substantial fraction of the highest energy cosmic rays are expected to be
photons in top-down cosmic ray models. In this  article, we consider the effect
that this will have on the UHECR spectrum observed by Auger. In particular, we
show that  Auger's data result in a spectrum without the appearance of a GZK
cutoff if the photon fraction of UHECRs follows the prediction of
top-down models.

\section{Top-Down Cosmic Ray Models}

If no GZK cutoff is found to be present in the UHECR spectrum, then either
local (within 10 to 50 Mpc) sources of UHECRs must exist \cite{cena}, or some
kind of exotic physics must be invoked to evade the GZK effect. Among exotic
possibilities, proposals have included UHECRs  composed of exotic hadrons
\cite{exotichadrons}, or strongly-interacting neutrinos \cite{exoticneutrinos},
or that protons can travel super-GZK distances due to a violation of Lorentz
invariance \cite{lorentz}.  The solution to the UHECR problem that we focus on
in  this article is a top-down scenario, in which the highest energy cosmic
rays are generated locally (in our galaxy) by the decays of supermassive
particles \cite{shdm} or topological defects \cite{topological}.
                                                                                                                             
\begin{figure}
\resizebox{8cm}{!}{\includegraphics{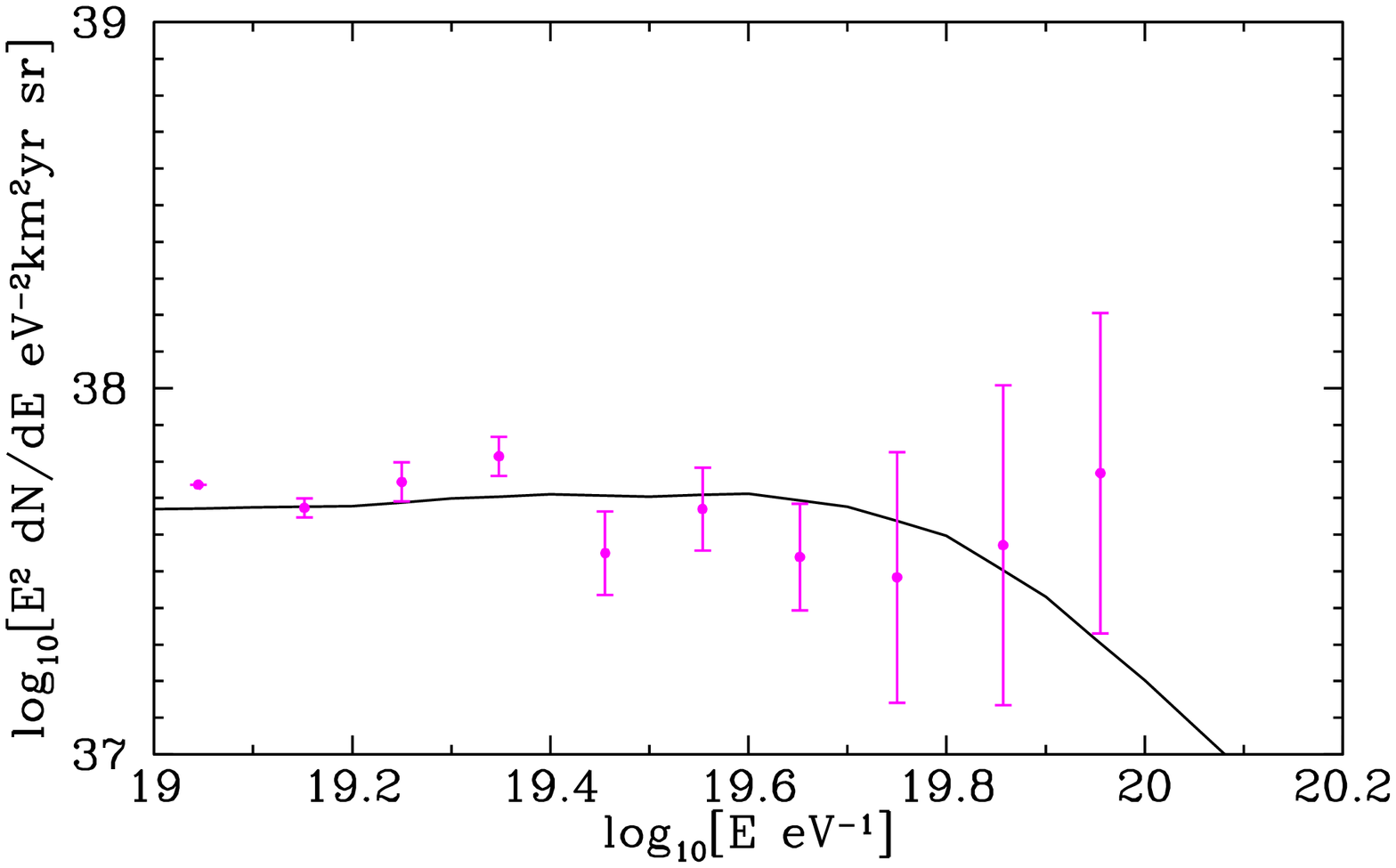}}
\resizebox{8cm}{!}{\includegraphics{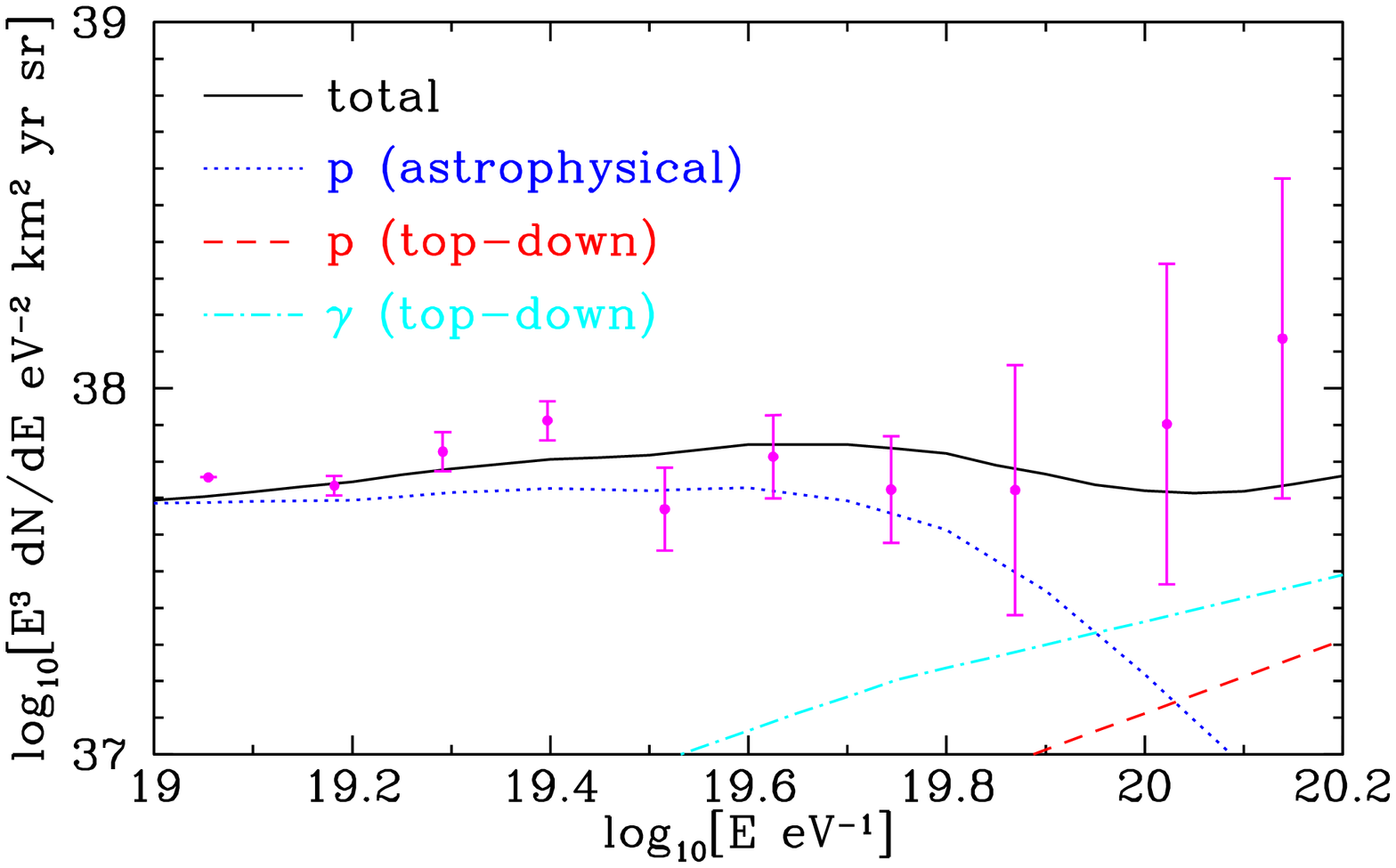}}
\caption{In the left frame, we plot the data from the Pierre Auger Observatory
compared to the spectrum from a conventional  astrophysical model of
homogeneously distributed sources with an injection spectrum of $dN_p/dE_p
\propto E_p^{-2.8}$.  In the right frame, we plot the same conventional source
spectrum along with the spectrum from the decay of supermassive  particles or
topological defects of mass $M_X = 6 \times 10^{21}$ eV. In the right frame,
the Auger data has been  shifted to account for the photon composition in
the top-down spectrum. Both models fit the data quite well.}
\label{spec}
\end{figure}

Unlike other proposed scenarios, in top-down models the highest energy cosmic rays are mostly
photons. For this reason, it can be misleading to compare
the spectrum presented by the Auger collaboration to the predicted  spectrum in
these models. In Fig.\ \ref{spec}, we compare the published Auger data to a
spectrum of protons from  homogeneous astrophysical sources (left frame) and to
the same spectrum plus a top-down component (right frame). We find, somewhat
unexpectedly, that the data  fits both scenarios reasonably well.
                                                                                                                             
In order to compare the Auger data with the predictions of a top-down model 
we consider the effect a proton-photon mixed primary composition would have on 
the Auger energy calibration curve. If at the present energy calibration range 
most of the primaries are protons, then the energy of a primary photon would be 
underestimated by a factor of two \cite{augerspec}. Consider a photon fraction 
$f(E,P_0,M_X)$, which is a function of the energy $E$,  the injection power $P_0$, and the 
mass of the decaying  particle $M_X$, which is small at the Auger energy calibration
range ($10^{18}\sim10^{19.4}$). Above this range a change in the calibration curve 
would lead to the following shift on the energy:
\begin{equation}
E_{shift}=E\left[1-f(E_{shift},P_0,M_X)+ 2 f(E_{shift},P_0,M_X)\right]
\end{equation}
where $E_{shift}$ is the shifted energy. For each bin of mean energy $E$ in the 
published Auger spectrum, we solved this equation to find the corresponding shifted 
energy $E_{shift}$.  We finally chose the values of $P_0$ and $M_X$ that best fit 
the data.

\begin{figure}
\resizebox{8cm}{!}{\includegraphics{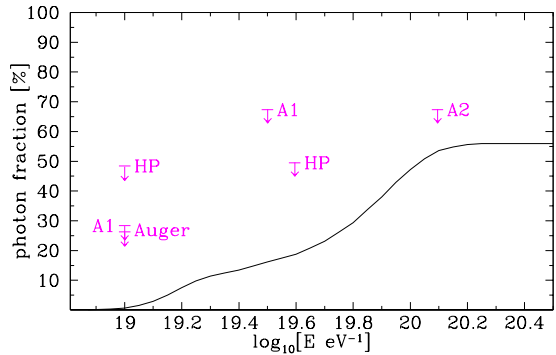}}
\caption{The fraction of photons in the UHECR spectrum as a function of energy
in the top-down model shown in the right frame of Fig.\ \ref{spec}. This
prediction is clearly below the limits set by the Auger \cite{augerphoton},
Haverah Park (HP) \cite{hpphoton} or AGASA  (A1, A2) data
\cite{agasaphoton}.}
\label{frac}
\end{figure}

The photon fraction of the UHECRs in the model shown in the right frame of
Fig.\ \ref{spec} is shown in Fig.\ \ref{frac}. The spectrum and composition of
UHECRs generated in top-down models has been calculated using the  publicly
available program SHDECAY \cite{tdspectra}. We have shown results here for the
case of decays to quark-antiquark pairs, and assumed the presence of
supersymmetry, although our conclusions are largely insensitive to  these
choices.

\section{Future Prospects For Confirming or Excluding Top-Down Models}

Although we have shown here that present UHECR data (Auger data in particular) 
is not capable of confirming or excluding top-down models, the prospects for 
testing such models in the near future are very promising. These prospects come 
from at least four types of observations: future improvement on the systematic 
uncertainty on the energy measurement, future UHECR photon fraction measurements, 
future UHECR anisotropy measurements, and future ultra-high energy neutrino 
measurements.

Auger will certainly improve its systematic uncertainties in the near 
future, leading to a better established energy spectrum. Such a spectrum would 
clearly resolve the issue of the existence of a GZK cutoff and confirm or rule out
the hypothesis of an additional high energy component.

As Auger accumulates more data, its hybrid detector will be able to place limits on
the photon fraction at increasingly high energies. Currently, the Auger photon 
limit only constrains the composition above $10^{19}$ eV to be less than 26\% photons
(at the 95\% confidence level) \cite{augerphoton}. To test most top-down scenarios, 
a similar constraint would be needed at a much higher energy, perhaps around $10^{19.7}$
eV. Auger will accumulate an exposure sufficient to accomplish this only after 
about 5--6 years of operation with a full southern array.

Auger will also be capable of studying the isotropy of the UHECR spectrum to 
unprecedented levels of precision. If a substantial fraction of the highest energy 
events are generated in top-down decays within the galactic halo, this can lead 
to an observable level of anisotropy directed toward the center of our galaxy. To 
identify such an isotropy, several hundred events of the highest energies will be 
required. It has been estimated that such a signal could be resolved at Auger South
after 3 years of operation with a full array~\cite{anisotropy}. Current Auger data 
only constrains anisotropies at 0.8--3.2 EeV \cite{anisotropydata}, well below the 
range effected by top-down models.

In addition to generating ultra-high energy photons and protons, top-down decays 
produce a large number of neutrinos. Such a flux of neutrinos is expected to result 
in of the order of one event in the first ANITA flight, scheduled for later in 
2006. Other experiments such as IceCube and Auger are expected to reach the 
sensitivity needed to detect top-down neutrinos as well~\cite{tdneutrinos}. Such neutrinos will be diffuse and difficult to identify as being of top-down origin, however.
The rates anticipated from ultra-high energy proton interactions with the cosmic 
microwave background (the cosmogenic neutrinos flux) are similar to those for 
top-down models, and are virtually impossible to distinguish from each other. The 
lack of such events, on the other hand, would be a fairly compelling piece of evidence 
against top-down models, and may imply a substantial component of heavy nuclei in the 
UHECR spectrum \cite{heavy}.

Among these four classes of observation, top-down models should testable within
the next several years, and are likely to be either experimentally excluded or confirmed.

\section{Summary and Conclusions}

The calibration of the Pierre Auger Observatory has been made using a hybrid technique 
at sub-GZK energies. This fact leads to a large systematic uncertainty at the highest 
energies of the order of about 40\%. A change in composition above Auger calibration 
energies might lead to a systematic shift in their higher energy events, even to the full extent
of this uncertainty. This is particularly true in top-down cosmic ray models, in which 
the highest energy cosmic rays are generated in the decays of super-massive particles or
 topological defects. In these models many of the highest energy cosmic rays are photons, 
which have their energies underestimated by about 50\% at Auger. 

In this article, we calculated the expected shift on the Auger spectrum assuming the 
photon content of a typical top-down model. This shift is consistent with the quoted 
experimental systematic uncertainty and is toward higher energies. We find that
the resulting spectrum agrees quite well to the top-down prediction. We 
also showed that the spectrum is consistent with a pure extragalactic proton 
hypothesis where no shift is needed.

As the Pierre Auger Observatory accumulates more data, its ability to calibrate in a
composition independent fashion will be applied at increasingly higher energies. At 
least 5--6 years of exposure with a full southern array will be required to reach the 
energy at which photons begin to dominate the UHECR spectrum in top-down models, 
however. Anisotropy measurements by Auger may also be able to test top-down models 
after a few years of observation, and upcoming ultra-high energy neutrino measurements
will be relevant to top-down models as well.

\medskip

{\it Acknowledgments:} We would like to thank Andrew Taylor for providing the conventional proton spectrum used in 
Fig.~\ref{spec}. We would also like to thank Aaron Chou for interesting discussions. This work has been supported by the US Department of Energy and by NASA grant NAG5-10842.

\end{document}